\RequirePackage{fix-cm}

\documentclass[twocolumn]{svjour3}          % twocolumn
\smartqed  % flush right qed marks, e.g. at end of proof
\usepackage{graphicx}
\usepackage{mathptmx}      % use Times fonts if available on your TeX system
\usepackage[english]{babel}
\usepackage{hyperref}
\hypersetup{
    colorlinks,
    citecolor=black,
    filecolor=black,
    linkcolor=black,
    urlcolor=black
}
\usepackage[font=small,labelfont=bf]{caption}
\usepackage{subcaption}
\newcommand*\samethanks[1][\value{footnote}]{\footnotemark[#1]}
\journalname{Journal of Quantitative Criminology}
\begin{document}

\title{Inference of the Russian Drug Community from One of the Largest Social Networks in the Russian Federation
}

\titlerunning{Inference of the Russian Drug Community from the Social Network LiveJournal}        % if too long for running head

\author{	L.J. Dijkstra 		\thanks{L.J. Dijkstra and A.V. Yakushev share first authorship of this work.}		\and
			A.V. Yakushev		\samethanks		\and
			P.A.C. Duijn		\and
			A.V. Boukhanovsky	\and
			P.M.A. Sloot
}


\institute{	L.J. Dijkstra \and P.M.A. Sloot \at
			Computational Science Department, University of Amsterdam (UvA), The Netherlands. \\ 
			\email{louisdijkstra@gmail.com}
			\and
			A.V. Yakushev \and L.J. Dijkstra \and A.V. Boukhanovsky \and P.M.A. Sloot \at
			High-Performance Computing Department, National Research University of Information Technologies, Mechanics and Optics (NRU ITMO), Saint Petersburg, Russia.\\ 
			\email{andrew.yakushev@yandex.ru}
			\and
			P.A.C. Duijn \at
			Criminal Intelligence Analysis, Dutch Police, The Netherlands. \\ 
			\email{pacduijn@gmail.com}
           \and
           A.V. Boukhanovsky \at
           \email{avb\_mail@mail.ru}
           \and
           P.M.A. Sloot (corresponding author) \at
           School of Computer Engineering (SCE), Nanyang Technological University (NTU), Singapore. \\ 
           Tel.: +31 (0) 20 525 7537\\
           Fax: +31 (0) 20 525 7419\\
           \email{p.m.a.sloot@uva.nl}
}

\date{10 May 2013}

\maketitle

\begin{abstract}
 
The criminal nature of narcotics complicates the direct assessment of a drug community, while having a good understanding of the type of people drawn or currently using drugs is vital for finding effective intervening strategies. Especially for the Russian Federation this is of immediate concern given the dramatic increase it has seen in drug abuse since the fall of the Soviet Union in the early nineties. 
Using unique data from the Russian social network `\emph{LiveJournal}' with over 39 million registered users worldwide, we were able for the first time to identify the on-line drug community by context sensitive text mining of the users' blogs using a dictionary of known drug-related official and `slang' terminology. By comparing the interests of the users that most actively spread information on narcotics over the network with the interests of the individuals outside the on-line drug community, we found that the `average' drug user in the Russian Federation is generally mostly interested in topics such as Russian rock, non-traditional medicine, UFOs, Buddhism, yoga and the occult. We identify three  distinct scale-free sub-networks of users which can be uniquely classified as being either `infectious', `susceptible' or `immune'.

\end{abstract}

\section{Introduction}
\label{sec:1}

Since the fall of the Soviet Union in the early nineties drug abuse has seen a dramatic increase in the Russian Federation. From 1990 to 2001 the number of registered drug addicts and drug-related crimes went up a nine- and fifteen-fold respectively (Sunami,  2007)\nocite{Sunami2007} and continued to rise over the last decade (Mityagin, 2012)\nocite{Mityagin2012}. The rapid spread and extent of this `drug epidemic' is of immediate concern to the Russian government and finding effective ways to halt this trend is considered to be of outmost importance. 

Due to the criminal nature and general social disapproval of drug use it is complicated to assess the drug community directly. Official governmental statistics do provide an insight into the general trend, but only manage to scratch the surface of the entire drug community in the Russian Federation. The drug users registered in their databases are often among the extreme cases: they have been in one (or more) rehabilitation programs or were arrested for using and/or selling illicit narcotics. The (still) `moderate' user stays out of the picture, making it difficult to obtain reliable information on the drug community as a whole. Within criminological research this non-registered crime is often referred to as \emph{dark number}, see Coleman and Moynihan (1996),\nocite{Coleman1996} and Rhodes et al. (2006)\nocite{Rhodes2006}.

Gaining a better understanding of what constitutes the drug community in the Russian Federation and in which ways its members can influence (or even inspire) others to start using might prove valuable for devising more effective intervening strategies that can turn the current situation for the better. 

In order to handle the drug society's inherent complexity, we will partition the Russian population into (roughly) three groups varying in their involvement in illicit drug use: 
\begin{enumerate}
	\item The \emph{immune}: the group of people that because of, for example, social commitments (e.g., marriage, children, job) and/or strongly held (religious) convictions will not be persuaded to start using drugs.
	\item The \emph{infectious}, i.e., the drug community: the group consisting of all individuals involved with drug abuse in one way or another (i.e., using, selling or producing).
	\item The \emph{susceptible} containing all individuals that are not a member of one of the previously mentioned groups. They are not involved in any way with illicit drug use at the moment, but might, due to their social position and environment be drawn toward drug use in the future.
\end{enumerate}

The idea to divide the population into these three groups was inspired by the division often used in models for virus spread, see for example the SIR-model of Daley and Kendall (1964)\nocite{Daley1964}, since a similar process seems to underlie the spread of drug addiction through society: infectious (drug users/ dealers) can infect susceptible others with the (drug) virus by means of direct and personal contact (i.e., sharing or selling drugs). This analogy has been made before, not only between virus spread and drug addition (Agar, 2005; Beenstock and Rahav, 2004; Mityagin, 2012)\nocite{Agar2005,Beenstock2004, Mityagin2012}, but also in the field of `obesity spreading' (Gallos et al. 2012) and for modeling the spread of information (Iribarren and Moro, 2009; Onnela et al., 2007; Bernardes et al., 2012)\nocite{Iribarren2009, Onnela2007, Bernardes2012}.

Social network sites (SNSs) have proved over the years that they provide means to uncover social structures and processes that were difficult to observe before (Scott, 2011)\nocite{Scott2011}. In this paper we investigate the social network site \emph{LiveJournal}\footnote{LiveJournal is available at \url{http://www.livejournal.com} (English) and \url{http://www.livejournal.ru} (Russian).}. With approximately 2.6 million registered Russian users and over 39 million registered users worldwide, it is one of the largest and most popular SNSs in the Russian Federation. The site offers its users an easy-to-use blog-platform where people can read and share their articles with others. In contrast to micro-blogging SNSs such as Facebook\footnote{Facebook is available at \url{http://www.facebook.com}.}  (Wilson et al., 2012; Ferri et al., 2012)\nocite{Wilson2012, Ferri2012} or Twitter\footnote{Twitter is available at \url{http://www.twitter.com}.} often mentioned in the literature, the site offers a tremendous amount of large user-written texts, making it extremely suitable for text-mining and, consequently, a unique source of data. Maybe because of having the impression to be among `friends', LiveJournal users write sometimes quite openly about their personal lives in their blogs. Some even comment on their use of drugs and their experiences with various kinds of narcotics. Others (the extreme cases) describe in detail the production process. These openly online expressions can be ascribed to the \emph{on-line disinhibition effect} (Suler, 2004)\nocite{Suler2004}; the invisible and anonymous qualities of on-line interaction lead to disinhibited, more intensive, self-disclosing and aggressive uses of language. Furthermore, recent studies show that criminal organizations are actively using on-line communities as a new `business' tool for communication, research, logistics, marketing, recruitment, distribution of drugs and monetarization (D\'ecary-H\'etu and \\ Morselli, 2011; \textsc{Europol}, 2011; Walsh, 2011; Choo and Smith, 2008; Williams, 2001)\nocite{Decary-Heta2011,Europol2011, Walsh2011, Choo2008, Williams2001}. Research of on-line communities, therefore, might aid in gaining a better understanding of the behavior of opaque networks within a society. 

In order to get a better insight into the drug community in the Russian Federation, we crawl a large randomly selected group of Russian LiveJournal users. Every blog entry of every user is associated with a weight indicating to what extent it refers to illicit forms of drug use by overlaying the document word-for-word with a dictionary consisting of known drug-related terminology (both official as well as informal/`slang'). When the sum of `indicator' weights of all the blog entries of a specific user reaches a certain threshold, the user is considered to be a member of the on-line drug community. The idea behind this approach is that drug users are more likely to use drug-related terminology in their blog entries than others. We will return to this assumption extensively  in Section 5. The way users are classified and the drug-dictionary are discussed in detail in Section 3.2. 

After identifying the on-line drug community, we might ask ourselves what kind of people are generally to be found in this sub-network? In order to get a better picture of the `average' user in this sub-community, we gather all the interests mentioned on each user's profile page and compare how often they appear within the on-line drug community with the frequency of appearance in the rest of the network. We limit ourselves here to interests, due to the fact that it is rather unclear how to automatically construct a `psychological profile' of a user based solely on his or her texts. That way, we try to isolate those interests that are truly more common in one of these two distinct groups of users. In Section 3.3 we describe the used methodology in more detail.

The susceptibility of people to the `drug virus' is thought to depend on their exposure to drug-related information and their own interest in this topic. This social mechanism of transmission is called \emph{differential association} in which drives, techniques, motives, rationalizations and attitudes toward deviant behavior are learned and exchanged by social interaction (Sutherland, 1947; Lanier and Henry, 1998; Haynie, 2002)\nocite{Sutherland1947, Lanier1998, Haynie2002}. From this perspective the number of interests a user has in common with the on-line drug community might indicate a higher susceptibility, since 1) this person is more likely to stumble upon blog entries published by member of the on-line drug community (which are more often about drug use), and 2) it might indicate a certain lifestyle more prone to drugs. Following this reasoning, we present a naive Bayesian classifier using the log-likelihood ratio method \\ (Kantardzic, 2011; Hastie et al., 2009)\nocite{Kantardzic2011, Hastie2009} in Section \ref{sec:3.4} that assesses the susceptibility of a user to drugs given his/her personal interests. When a user's interests overlap more with the interests in the on-line drug community than the interests of the rest of the population, they are considered to be susceptible.

Users that were not identified as being a member of the on-line drug community on the basis of their written texts or as susceptible due to a large similarity with their interests and the interests common in the on-line drug community are considered to be immune. They do not write (much) about illicit drug use and their interests do not suggest a lean towards the on-line drug community.

After having (roughly) identified the three subgroups (i.e., immune, infectious and susceptible) in the social network LiveJournal, we might wonder whether there are structural differences between the corresponding subnetworks. In Section \ref{sec:4.3} we will describe and compare them. 

The remainder of this paper is organized as follows. In Section \ref{sec:2} we discuss the social network site LiveJournal, describe the kind of information users put out about themselves and point to several unique features this SNS has over others often studied in the literature. Section \ref{sec:3} describes the crawled LiveJournal data set and the methods used to partition its users and determine significant interests. The results are presented in Section \ref{sec:4}. We will finish with several conclusions, a rather extensive discussion and a few pointers for future research. In Appendix 1 we explore the frequency with which interests appear in the network and show that this probability distribution follows a power-law. 

\section{The SNS LiveJournal}
\label{sec:2}

The social network site LiveJournal with over 39 million worldwide and approximately 2.6 million registered Russian users is by far the most popular blog-platform in the Russian Federation. With 1.7 million active users and (approximately) 130,000 new posts every day the site offers a fast body of data for studying social structures and processes\footnote{LiveJournal's own statistics page can be found at \url{http://www.livejournal.com/stats.bml}.}. In addition to publishing their own articles, the users are offered the possibility to enter information on their whereabouts (e.g. hometown), demographics (e.g. birthday), their personal interests (e.g. favourite books, films and music) and even their current mood (e.g. happy, sad). Articles can be tagged and an extensive comment system provides the readers with the possibility to respond and exchange opinions and ideas.  

Users can unilaterally declare any other registered user as a `friend', i.e., ties are unidirectional. A tie reflects the desire of a user to keep up-to-date with the articles of the other. Consequently, every profile contains two lists of ties: 1) a list of alters that currently follow the articles published by the ego, and 2) a list of alters whose articles the ego follows. (Note the similarity with Twitter). We will refer to these lists as the list of \emph{followers} and \emph{following friends}, respectively. 

LiveJournal differs from other (large) social network sites in two important aspects: 1) it has a large number of users that actively write in Russian, and 2) the texts are large in contrast to the micro-blogging SNSs often considered in the literature (Wilson et al., 2012)\nocite{Wilson2012}. The latter makes LiveJournal exceptionally suitable for text-mining and, as such might provide insights into social structures and processes where other SNSs cannot.

\section{Methods}
\label{sec:3}

Section \ref{sec:3.1} describes the data collected from the SNS LiveJournal. In Section \ref{sec:3.2} we discuss the drug-dictionary and procedure used for classifying those users who are most likely to be involved in drug abuse. After colouring the subnetwork of the on-line drug community, we proceed in Section \ref{sec:3.3} with identifying those interests that are more common for this set of users or the rest of the on-line. These indicative interests are used by the naïve Bayesian classifier introduced in Section \ref{sec:3.4} for identifying the `susceptible' and `immune' subnetworks. We will later analyze the structure of these three subnetworks later in Section \ref{sec:4.3}.

\subsection{The LiveJournal Data Set}
\label{sec:3.1}

On the 9th of September 2012 we crawled 98602 randomly selected Russian user profiles. For each profile we stored its username, the last 25 posted blog entries, personal interests and the lists of followers and following `friends'. In addition we stored (when available) the user's birthday and place of living.

In order to collect this data, we developed a distributed crawler that employs the MapReduce Model (L\"ammel, 2007)\nocite{Lammel2007} and the open source framework Apache Hadoop (White, 2009)\nocite{White2009}. The system is similar to the Apache Nutch crawler (Cafarella and Cutting, 2004)\nocite{Cafarella2004} but allows for multiple users to collect and process data at the same time; the fetcher module is moved outside the Hadoop framework making it a separate application that can run on various machine architectures simultaneously.

A total of 22357 users fully specified their birthday on their public profile (ages higher than 80 were regarded to be reported falsely). In Section 4.1 we explore some characteristics of the crawled population and compare it with the Russian population. 

\subsection{The On-line Drug Community}
\label{sec:3.2}

Users are classified as being a member of the on-line drug community by comparing their last 25 blog entries with a dictionary of known drug-related terminology collected by drug experts at the Saint Petersburg Information and Analytical Center\footnote{The homepage of SPb IAC can be found at \url{http://iac.spb.ru} (in Russian).}. The total of 368 words in this dictionary are split up into two categories: official and informal/`slang' terminology. Official terminology are words that are unmistakingly related to illicit drug use (e.g., cocaine and heroin) and are assigned a high weight, i.e., 5. Informal/`slang' expressions can often be interpreted in various ways and cannot be directly related to drug use. For example, the Russian word `kolesa' refers normally to wheels while it also can be used (in rather dubious circumstances) as a word for pills. To account for this ambiguity, `slang' expressions are assigned a lower weight than official terminology, i.e., 1. Table \ref{tab:1} shows a few example words from the dictionary alongside their weight and (free) English translation\footnote{The full drug-dictionary is freely available and can be downloaded at \url{http://escience.ifmo.ru/?ws=sub48}.}.

In addition to this set of words, each blog entry was also checked for the presence of a collection of drug-related phrases. The presence of certain combinations of words in a text, e.g., `injecting' and `heroin', is a strong indication that the author is involved with illicit drug use. In order to account for this valuable information, the dictionary consists additionally of 8359 phrases, each assigned with a slightly higher weight than the mere sum of the words it consists of\footnote{The number of phrases (8359) is rather high in comparison to the number of words (368) in this dictionary. This is due to the fact that we consider a phrase consisting, for example, of the words `injecting', `heroin' and the phrase with the words `injection', `heroin' and `needle' as two separate expressions (where the latter is associated with a higher weight than the former).}.   

In order to compare inflected or derived words in the posts with words in the dictionary we first reduce them to their root form using a Russian version of the Porter stemming algorithm (Porter, 1980; Porter, 2006)\nocite{Porter1980,Porter2006}. 

 When the summed weights of all the blog entries of a user reaches a certain threshold, he/she is considered to be a member of the on-line drug community.  Users who use a small number of the words and phrases from the dictionary in a limited number of blog entries are, thus, less likely to be identified as a member than the ones who frequently use drug-related terminology throughout a large numbers of texts. The threshold was set manually, see Fig. \ref{fig:1}. 

We will refer to the entire set of users who's summed weights reaches the threshold as the on-line drug community throughout the rest of this paper. To what extent the sub-community corresponds to the Russian drug community will be a point of discussion in Section 5. 

\begin{table}[h!]
\centering
\caption{Examples of words in the drug-dictionary}
\label{tab:1}      
	\begin{tabular}{lcc}
		\hline\noalign{\smallskip}
		Russian 	& English translation 		& Weight  \\
		\noalign{\smallskip}\hline\noalign{\smallskip}
		Kokain		&	Cocaine					& 5 \\
		Geroin		&	Heroin					& 5 \\ 
		Mariguana	&	Marijuana				& 5 \\ 
		Abstyag		& 	Withdrawal syndrome		& 5 \\ 
		Tabletki	& 	Pills					& 1 \\ 
		Kolesa		&	Pills/Wheels			& 1 \\ 
		\noalign{\smallskip}\hline
	\end{tabular}
\end{table}

\subsection{Identifying Common Interests of the On-line Drug Community}
\label{sec:3.3}

In this section we will formulate an approach for determining which interests are most common (or uncommon) for a particular subset of SNS users, in our particular case, the on-line drug community. 

First, we collect the interests on the profile pages of all users in the on-line drug community that at least appear more than 10 times. (The reason for disregarding rather unfrequent interests is that they do not add much when one wants to gain a better understanding of an entire community). Lets denote this set of interests with $\mathbf{I} = \{I_1, I_2, \dots, I_m\}$. Since the members of the on-line drug community are known, we are able to count how often users express their interest in both this sub-community and the rest of the social network. For every interest $I_i$ we can, thus, obtain a $2 \times 2$ contingency table similar to Table \ref{tab:2} 
\begin{table}[h!]
\centering
\caption{The $2 \times 2$ contingency table for interest $I_i$}
\label{tab:2}       
\begin{tabular}{l | c c | c}
 & Drug community & Rest & \emph{Total} \\ \hline
Is interested in $I_i$ & $a$ & $b$ & $a + b$ \\   
Not interested in $I_i$ & $c$ & $d$ & $c + d$ \\ \hline
\emph{Total} & $a + c$ & $b + d$ & $n$ 
\end{tabular}
\end{table}
where $(a + b + c + d) = n$ is the total number of users in the crawled population that have at least one interest on their profile page (i.e., $n = 62370$), $a + c$ is the number of users identified as members of the on-line drug community, $a + b$ is the total number of users who expressed their interest in $I_i$ and $c + d$ are the users not interested in $I_i$. The question is whether this interest appears significantly more (or less) in the on-line drug community than in the rest of the rest of the network, i.e., do the proportions $a/(a + c)$ and $b/(b + d)$ differ?

We, thus, have $m$ null hypotheses ($H_i^0$), one for each interest $I_i$ in $\mathbf{I}$. Applying the two-sided version of Fisher's exact test\footnote{A $\chi^2$ test originally designed for $2 \times 2$ contingency tables by Sir R.A. Fisher (1922).} (Fisher, 1922; Agresti, 1992)\nocite{Fisher1922, Agresti1992} to each contingency table provides us with their corresponding $p$-values: $p_1, p_2, \\ \dots, p_m$. 

The total number of null hypotheses is large (3282 to be precise, corresponding to the total number of interests expressed more than 10 times in the on-line drug community). Simply comparing the obtained $p$-values with a common fixed significance level (e.g., $p \leq .05$) will result in a high number of false discoveries, i.e., falsely rejected null hypotheses. Benjamini and Hochberg (1995)\nocite{Benjamini1995} showed that the expected false discovery rate can be upper bounded by $q \in [0,1]$ with the following control procedure\footnote{Strictly speaking, the expected false discovery rate is only upper bounded when the $m$ test statistics are independent, which does not hold in this particular case. B. Efron makes the case in his book \emph{Large-Scale Inference} (2010)\nocite{Efron2010} that this independency constraint is not strong.} (Benjamini and Hochberg, 1995; Benjamini and Yekutieli, 2001)\nocite{Benjamini1995,Benjamini2001}: 
\begin{enumerate}
	\item Order the $p$-values in increasing order, i.e., $p_{(1)} \leq p_{(2)} \leq \dots \leq p_{(m)}$. 
	\item For a given $q$, find the largest $k$ for which $p_{(k)} \leq kq$.
	\item Reject all $H^0_{(i)}$ for $i = 1,2,\dots,k$.
\end{enumerate}
We will use a $q$-value of $5\%$. The interests associated with all rejected $H^0_{(i)}$, 
$\mathbf{I'} = \left\{ I_{(1)}, I_{(2)}, \dots, I_{(k)} \right\}$, are considered to be the interests that really differ between the on-line drug community and the rest of the social network. 

Due to the large sample size and the initially large number of interests, the number of significant interests in $\mathbf{I'}$ is expected to be quite high. Partitioning them into a set of \emph{themes} might help with getting a better overview of the wide variety of significant interests. In order to do so, we cluster the set of significant interests $\mathbf{I'}$ using a hierarchical agglomerative clustering algorithm with a complete linkage strategy (Kantardzic, 2011; Everitt, 2001)\nocite{Kantardzic2011, Everitt2001}. Complete-linkage is preferred here over single-linkage due to the fact is does not suffer from the chaining phenomena, i.e., clusters may be forced together due to single elements being close to each other, even if a majority of elements is very distant. Average-linkage was no option due to its high computational load. The similarity between two clusters of interests, $C_1$ and $C_2$, is defined as 
\begin{equation}
	sim(C_1, C_2) = \frac{n(S_1 \cap S_2)}{\sqrt{n(S_1) \cdot n(S_2)}}
	\label{eq:1}
\end{equation}
where $S_1$ and $S_2$ are the sets of users that expressed their interest in at least one of the topics in, respectively, $C_1$ and $C_2$.  $n(\cdot)$ returns the number of users. This similarity measure is known as \emph{cosine similarity} or more commonly known in biology as the Ochiai coefficient (Ochiai, 1957)\nocite{Ochiai1957}. We will refer to the resulting clusters of significant interests as themes throughout the rest of this paper.

\subsection{Assessing Susceptibility} 
\label{sec:3.4}

A large number of common interests between a user and the on-line drug community might indicate a higher susceptibility to drugs, since 1) the user is more likely to stumble upon blog entries published by members of this sub-community, and 2) it might indicate a certain lifestyle more prone to drug use. Certain interests might, on the other hand, indicate a low susceptibility. Think of interests that suggest that the user in question has certain social commitments (e.g., marriage, children, job) or strong-held (religious) convictions. The idea that interests are related to susceptibility underlies the classification method in this section: an individual is considered to be a susceptible user when his/her personal interests resemble the interests common for the drug community more than the interests of the rest of the on-line population. 

A naive Bayesian classifier was used (Kantardzic, 2011)\nocite{Kantardzic2011}. Due to the fact that certain combinations of interests are rare, we are forced to assume conditional independence between each pair of interests and use the log-likelihood ratio method.  

Let us first define $k$ feature variables, one for each interest in the set $\mathbf{I'}$: 
$$
	\mathbf{F} = \{F_1, F_2 ,..., F_k\} 
$$
where $F_i$ is true when the user is interested in $I_i$ in $\mathbf{I'}$ and otherwise false. The set of feature variables $\mathbf{F}$ is used to describe the personal interests of each user in the network.

The chance that a user belongs to the drug community ($D$) given his/her interests is given by the conditional chance $P(D \mid \mathbf{F})$. Given the assumption that each feature variable $F_i$ is conditionally independent of $F_j$ when $i \neq j$, i.e., $P(F_i \mid D, F_j) = P(F_i \mid D)$, this probability can be expressed as 
\begin{equation}
	P (D \mid \mathbf{F}) = \frac{P(D)}{P(\mathbf{F})} \displaystyle\prod_{i = 1}^k P(F_i \mid D).
	\label{eq:2}
\end{equation}
Similarly, the chance of not being a member of the drug community given the users interests is
\begin{equation}
	P (\neg D \mid \mathbf{F}) = \frac{P(\neg D)}{P(\mathbf{F})} \displaystyle\prod_{i = 1}^k P(F_i \mid \neg D).
	\label{eq:3}
\end{equation}
By applying the log-likelihood ratio method, i.e., dividing eq. (\ref{eq:2}) by eq. (\ref{eq:3}) and taking the natural logarithm of both sides, we find that the inequality $P(D \mid \mathbf{F}) > P(\neg D \mid \mathbf{F})$, i.e., the user is more likely to belong to the drug community given the user's interests, is equivalent to the inequality: 
\begin{equation}
	\log \frac{P(D)}{P(\neg D)} + \displaystyle\sum_{i=1}^k \log \frac{P(F_i \mid D)}{P(F_i \mid \neg D)} > 0. 
	\label{eq:4}
\end{equation}
A user is considered to be susceptible when he/she does not belong to the drug community and this inequality holds. Users that are not a member of the on-line drug community or considered to be susceptible, are immune.

\section{Results}
\label{sec:4}

\begin{figure}
	\centering
  	\includegraphics{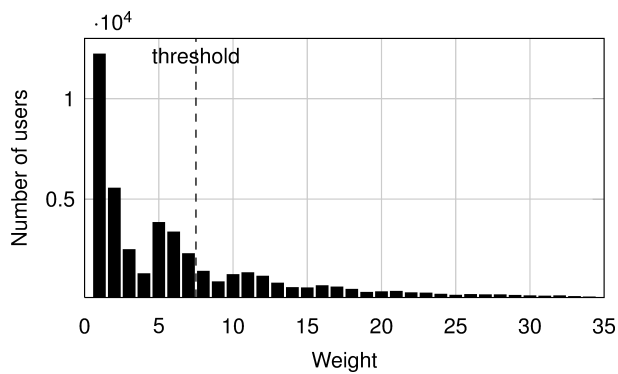}
	\caption{The summed weights of the blog entries of each user in the LiveJournal data set. The higher the summed weight the more the user used the words and phrases present in the drug-dictionary (see Section \ref{sec:3.2}). Users are considered to be a member of the on-line drug community when their weighted sum crosses the threshold of $8$}
	\label{fig:1}    
\end{figure}

In order to identify those users in the network involved with illicit drug use, we overlaid their last 25 blog entries with a dictionary of known drug-related terminology (see Section \ref{sec:3.2}). Fig. \ref{fig:1} depicts the distribution of the weights assigned to the randomly crawled LiveJournal users. Note that the majority of users appear to make use of a rather small number of drug-related terminology. The fluctuations that can be seen around the weights 5, 10 and (less distinct) 15 and 20 can be explained by the weights assigned to the words present in the drug-dictionary (5 for official, clearly drug-related, terminology and 1 for (ambiguous) `slang' expressions). The users with the highest weights are assumed to be the ones most interested and/or involved in illicit drug use. The threshold was set to 8 (see Fig. \ref{fig:1}), i.e., when the weight of a user crosses 8, he/she is considered to be a member of the on-line drug community. Other thresholds close to 8 were considered as well. We found that the themes as presented in Section \ref{sec:4.2} did not change tremendously. By setting the threshold to 8, approximately 20\% of the total set of crawled users were classified as being a member of the on-line drug community.

\subsection{Characteristics of the SNS LiveJournal}
\label{sec:4.1}

\begin{figure*}[hbt]
        \centering
        \begin{subfigure}[b]{0.49\textwidth}
                \centering
                \includegraphics[width=\textwidth]{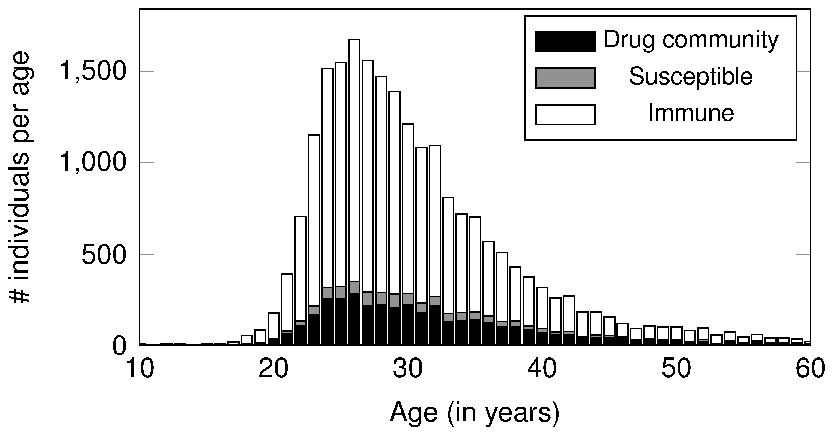}
                \caption{LiveJournal (2012)}
                \label{fig:2a}
        \end{subfigure}
        ~ 
        \begin{subfigure}[b]{0.49\textwidth}
                \centering
                \includegraphics[width=\textwidth]{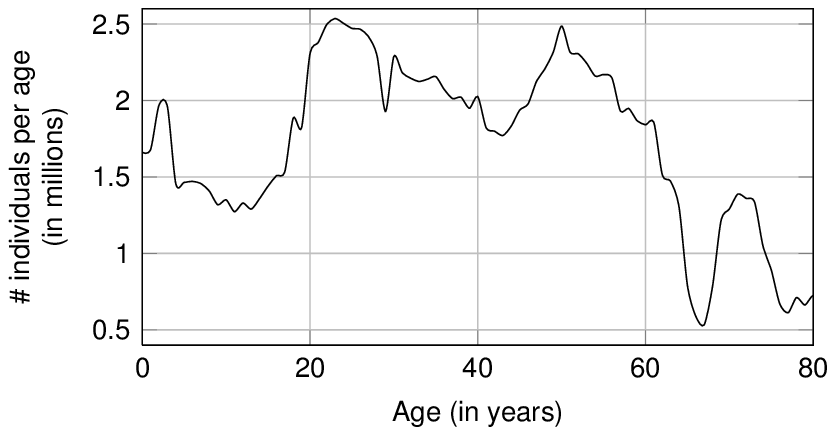}
                \caption{Russian Federation (2011)}
                \label{fig:2b}
        \end{subfigure}
        \caption{\textbf{a} The age distribution of the LiveJournal data set (2012) split out between the on-line drug community, and the susceptible and immune user groups. Note that this SNS is especially popular in Russia among 20 to 40 year olds \textbf{b} The age distribution of the Russian Federation on the 1st of January 2011 (the data was made available by Rosstat). Note the difference between the two age distributions. LiveJournal does, thus, not provide a good sample of the Russian population, although, while investigating illicit drug use it might be useful to sample especially that fraction of the population known to be most involved with narcotics (Mityagin 2012)}
        \label{fig:2}
\end{figure*}

Fig. \ref{fig:2}a depicts the age distribution of the LiveJournal data set split out between the on-line drug community and the susceptible and immune user groups. Note that this SNS is especially popular among 20 to 40 year old individuals. Figure \ref{fig:2}b depicts the age distribution of the Russian Federation as determined on the 1st of January 2011. The data was made available by Rosstat\footnote{The governmental statistics agency of the Russian Federation. They can be found at \url{http://www.gks.ru} (in Russian) with links to their rather extensive database.}. The major dip around the ages 62-70 is a reflection of the impact that the Second World War had on the Russian population. 

Note the difference between the Russian LiveJournal community and the Russian population as a whole. Using LiveJournal to sample the Russian population poses two problems: 1) one only samples those individuals who are registered as a user in this SNS, and 2) we seriously oversample the age group 20-40. Both aspects might not pose a real threat; the Russian drug community is, as mentioned before, difficult (or even impossible) to sample directly, making sampling a SNS one of the limited options one has, when one wants to gain a better insight into this sub-community. In addition, illicit drug use is known to occur especially in this particular age group (Mityagin, 2012)\nocite{Mityagin2012}. The strong presence of this group, thus, might help in gathering more information on the community of interest.    

Of the total number of 98602 users studied in the LiveJournal data set, 16553 and 3586 were identified as, respectively, members of the drug community and susceptible users. Susceptible users are identified using the naive Bayesian classifier as described in Section \ref{sec:3.4} which makes use of the interests the user posted on his/her profile page. Common interests can be shown to be rare. In fact, the frequency with which an interest is mentioned by users of this SNS can be shown to follow a power-law distribution with coefficient $\gamma \approx 1.54$, see Appendix 1. With a low number of common interests, there is often not enough to go on in order to reliably classify a user as being susceptible, which explains the relatively small number of susceptible users found.

\subsection{Drug Indicators}
\label{sec:4.2}

\begin{figure*}[t]
	\centering
	\includegraphics{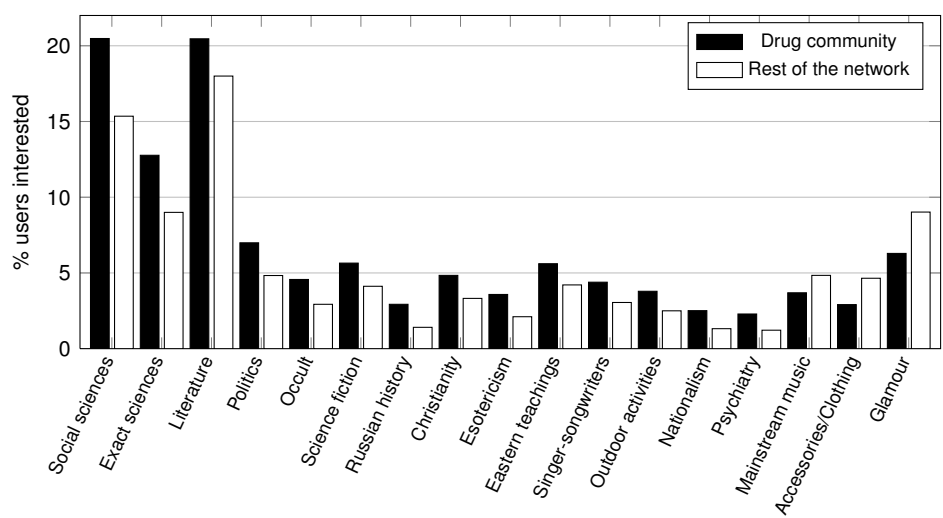}
	\caption{The percentage of users within the on-line drug community and the rest of the on-line population interested in each theme (see Table \ref{tab:3}). Users are considered to be interested in a theme when they mention at least one of the interests contained in that theme. Note that the last three themes are more likely to be found in the non-drug section of the network. The other themes are relatively more likely to appear in the on-line drug community}
	\label{fig:3}
\end{figure*}

After applying Fisher's exact test and Benjamini and Hochberg's false discovery rate control procedure with a $q$-value of 5\% (see Section \ref{sec:3.3}), we found 268 of the 3282 initial interests to be significant, i.e., the on-line drug community is, thus, more/less interested in these topics than the rest of the LiveJournal users. In order to assess to what extent an interest $I$ is indicative for being a member of the drug community ($D$) or the rest of the population, we use the conditional probability $P(D \mid I)$. Among the interests most indicative for the on-line drug community (i.e., $P(D \mid I) > .5$), we found interests such as: the White Movement (a loose confederation of anti-communist forces who fought the Bolsheviks in the Russian civil war; now often associated with the Russian nationalistic movement), humanistic psychology, partisans, Aryan (ancient people that partly inhabited current Russian territory), Stalinism, Dadaism, narcology and Magadan (a city in the far east of the Russian territory, famous for its large jail). Among interests most indicative for not belonging to the drug community (i.e., $P(D \mid I) < .5)$, we found interests such as: accessories, beads, jewellery, London, clothing, glamour, handmade, shoes, beach and interior design. 

In order to get a better view on the wide variety of significant indicative interests, we clustered them using the cluster algorithm described in Section \ref{sec:3.3}. We found 42 different themes in total. In this Section we will only discuss the ones most prominent within the on-line drug community and the rest of the LiveJournal population. 

Fig. \ref{fig:3} shows the various themes and to what extent they appear in the on-line drug community and the rest of the LiveJournal population. We consider users to be interested in a theme, when they mention at least one of the interests contained in that theme on their profile page. 

The names assigned to each theme were determined by the writers of this article. In order to overcome some of the inevitable subjectivity inherent to this process, we will describe the themes shortly in Table \ref{tab:3}, where the second column denotes the number of significant interests in each theme. When the number of interests in a theme is small, we will sum up all the interests (translated to English); otherwise we will suffice with a short description. 

In both Fig. \ref{fig:3} and Table \ref{tab:3} the last three themes (Mainstream music, Accessories/Clothing and Glamour) appear more often in the non-drug section of the network. The others are more common for the on-line drug community. 

Recall that significant interests were clustered solely on the basis of their cosine similarity (i.e., the more users that expressed their interest in both topics, the higher the `similarity'). In which of the two distinct communities the interest is more prominent is not taken into account. Each theme is, thus, likely to contain interests that are more common for the on-line drug community and interests that are more often found in the rest of the network. To what extent a theme can be related to one of these two groups can, therefore, be expected to be less clear than for individual interests.

\begin{table*}[hbt]
	\centering
	\caption{Description of the most prominent themes}
	\label{tab:3}  
	    
\begin{tabular}{l l p{.65\textwidth}}
	\hline\noalign{\smallskip}
	Theme & \# & Description  \\
	\noalign{\smallskip}\hline\noalign{\smallskip}
	Social sciences & $5$ & Sociology, history, economics, psychology and law. \\
	Exact sciences & $7$ & Programming, biology, astronomy, medicine, archeology, ecology and philosophy. \\ 
	Literature & $9$ & Containing rather general interests such as books, journalism, poetry and prose. \\
	Politics & $22$ & This theme contains various national (opposition, corruption and Russia), international (Chechnya, NATO, Poland and Ukraine) and general (socialism, democracy and anti-communism) political topics. \\ 
	Occult & $15$ & Concerns a wide variety of topics, including, for example, the occult, non-traditional medicine, mysticism, clairvoyance, telepathy and the prediction of the future through the reading of cards (tarot). \\
	Science fiction & $8$ & Containing interests like UFOs, futurology, nanotechnology, science fiction and the American science fiction writer H. Harrison. \\ 
	Russian history & $11$ & Ranging from general sciences (anthropology, ethnography, war history) to particular events in the history of Russia (WWII, the Russian civil war) and important historical groups (partizans). \\
	Christianity & $3$ & God, the Russian orthodox church and religion.\\
	Esotericism & $7$ & Contains various topics related to esotericism (esotericism itself, but also the expansion and altering of the human mind) and Castaneda, a rather famous author who popularized topics such as `stalking' (technique to control the mind) and lucid dreams. \\
	Eastern teachings & $10$ & Various eastern teachings/religions (Buddhism, Zen and yoga) and related terms (e.g., mantras, chakras and tantras). \\ 
	Singer-songwriters &  $6$ & Interests related to Russian rock and singer-songwriters (e.g., V. Vysotsky). \\
	Outdoor activities & $8$ & Diving, fishing, hunting and topics related to Mountain climbing (e.g., alpinism and the Altai mountains) and survival. \\
	Nationalism & $9$ & Covering interests such as the Russian empire, patriotism, the Russian people, the White Movement and antiglobalization. \\
	Psychiatry & $6$ & Including psychiatry, psychoanalysis, psychotherapy, psychosomatic medicine and transpersonal and humanitarian psychology. \\
	Mainstream music & $6$ & Containing several famous mainstream musicians, such as Madonna, Coldplay and Bj\"ork. \\
	Accessories/Clothing & $13$ & Varying from accessories like beads, jewelry, shoes and bags to clothing and interior design. \\
	Glamour & $13$ & Includes the interest glamour itself. It further covers fashion (e.g., journals, style, jeans, design and shopping) and the night-life of Moscow. \\ 
	\noalign{\smallskip}\hline
	\end{tabular}
\end{table*}

\subsection{Network Structure Analysis}
\label{sec:4.3}

In this section we will explore the structure of the on-line drug community, susceptible and immune subnetworks. 

\begin{figure*}[hbt]
        \centering
        \begin{subfigure}[b]{0.544\textwidth}
                \centering
                \includegraphics[width=\textwidth]{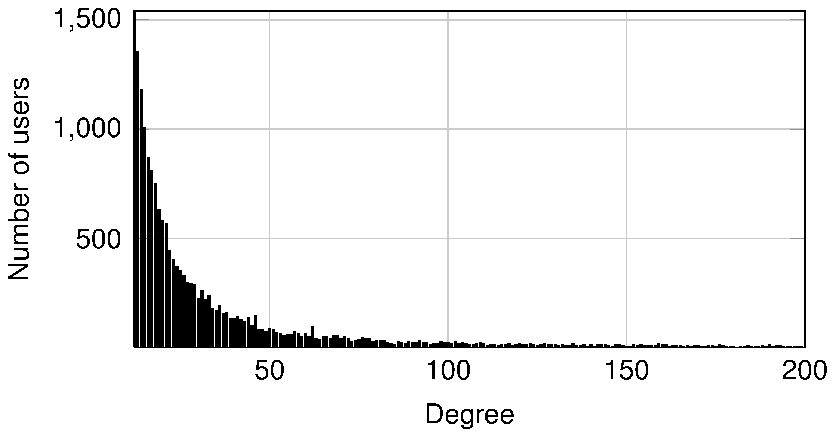}
                \caption{}
                \label{fig:4a}
        \end{subfigure}%
        ~ 
        \begin{subfigure}[b]{0.232\textwidth}
                \centering
                \includegraphics[width=\textwidth]{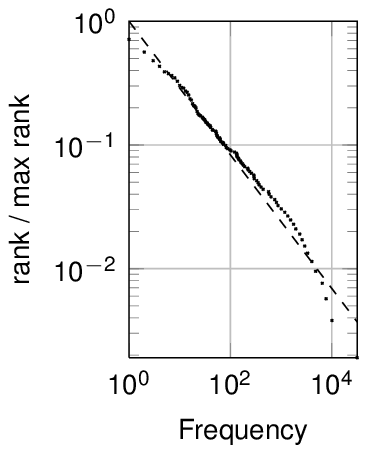}
                \caption{}
                \label{fig:4b}
        \end{subfigure}
        \caption{\textbf{a} A fraction of the degree distribution of the crawled LiveJournal network. Note that the number of users decreases exponentially with degree \textbf{b} The rank/frequency log-log plot of the degree distribution and the power-law fit depicted as a dashed line ($\gamma \approx 1.54$ and $x_{min} = 8$). The $p$-value was found to be approximately $.57$, i.e., there is no reason to believe that the degree distribution does not follow a power-law}
        \label{fig:4}
\end{figure*}

Fig. \ref{fig:4}a shows the degree distribution of the total crawled LiveJournal network. Degree is defined here as the number of followers and following `friends' of a user. Note that the number of users seems to decrease exponentially with degree; an indication that the distribution might follow a power-law:
\begin{equation}
	p(x) = Cx^{-\gamma}
	\label{eq:5}
\end{equation} 
where $x$ is the degree of a user, $\gamma$ is the power-law coefficient and $C$ is a constant. Power-law distributions appear in a wide variety of natural and man-made processes, e.g., the number of inhabitants in cities, the diameter of moon crates and the intensity of solar flares. The wide-spread appearance of the power-law raises the question whether the same process might underlie these (at first glance) different phenomena, causing quite a discussion in the literature. For a more elaborate discussion of power-laws and their appearance, we refer the reader to a recent paper by Pinto et al. (2012)\nocite{Pinto2012}. 

Fig. \ref{fig:4}b shows the rank/frequency log-log plot\footnote{A rank/frequency log-log plot is the plot of the occurrence frequency versus the rank on logarithmically scaled axes. For a more elaborate description on how to construct such a plot, see the paper by Mark Newman (2005)\nocite{Newman2005}, Appendix A.} of the degree distribution in \ref{fig:4}a. Note the points in this plot lie (approximately) on a straight line, which is a characteristic of power-law distributions. 

Very few real-word networks display a power-law distribution over the entire degree range, making it necessary to determine where the degree distribution is most likely to start following a power-law (denoted here with $x_{min}$). The power-law exponent $\gamma$ and $x_{min}$ were determined using the maximum likelihood method as described in the paper by Clauset et al. (2009)\nocite{Clauset2009} and were found to be equal to 1.54 and 8, respectively. The fit is shown in Fig. \ref{fig:4}b as a dashed line. Note that the line seems to fit the data quite well. The standard statistical test for the quality of fit as proposed by A. Clauset, C. Shalizi and M. Newman (2009)\nocite{Clauset2009} shows that the data gives no raise to believe that the degree distribution does not follow a power-law (i.e., $p = .57$ with 1000 repetitions). 

\begin{figure*}[hbt]
        \centering
        \begin{subfigure}[b]{0.32\textwidth}
                \centering
                \includegraphics{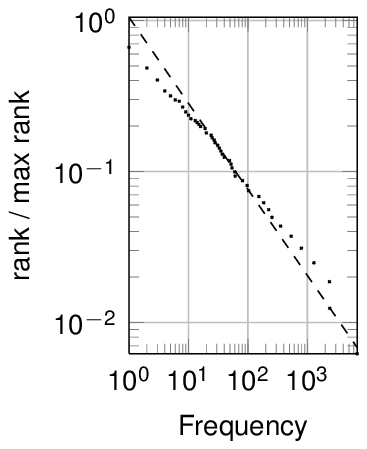}
                \caption{}
                \label{fig:5a}
        \end{subfigure}%
        ~
        \begin{subfigure}[b]{0.32\textwidth}
                \centering
                \includegraphics{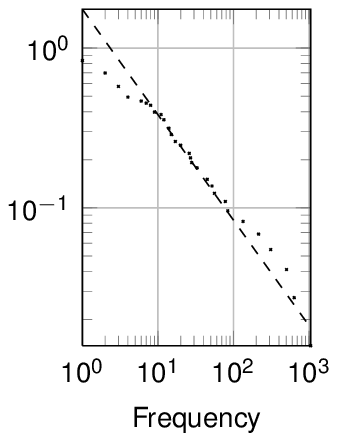}
                \caption{}
                \label{fig:5b}
        \end{subfigure}
        ~
         \begin{subfigure}[b]{0.32\textwidth}
                \centering
                \includegraphics{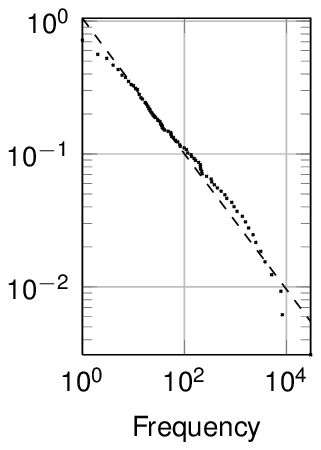}
                \caption{}
                \label{fig:5c}
        \end{subfigure}
        \caption{The rank/frequency log-log plots of the degree distributions of the three subnetworks in the crawled LiveJournal network: the drug community ($\gamma \approx 1.57$  and $x_{min} = 19$) and the susceptible ($\gamma \approx 1.66$ and $x_{min} = 8$) and immune subnetwork ($\gamma \approx 1.54$ and $x_{min} = 10$) . The power-law fit is depicted as a dashed line. The found $p$-values give no reason to believe these distributions do not follow a power-law}
        \label{fig:5}
\end{figure*}

Fig. \ref{fig:5} shows the rank/frequency log-log plots of the degree distributions of the on-line drug community, susceptible and immune network together with their power-law fits. Note that these sub-networks also follows a power-law distribution, only with slightly different $\gamma$'s.

Table \ref{tab:4} shows various characteristics of the LiveJournal network and its three subnetworks. Standard deviations are reported between parentheses. Note that the mean age does not differ much. The large differences between the maximum degrees of these networks are common for heavy right-tailed distributions. The best fits for $\gamma$, $x_{min}$ and the $p$-value of the goodness of fit test are reported as well.

\begin{table*}[hbt]
	\centering
	\caption{Structural characteristics of the various subnetworks in LiveJournal}
	\label{tab:4}  
	\begin{tabular}{l c c c c c c c c} 
		\hline\noalign{\smallskip}
		Network & Size & Edges & Age & Max. degree & $\gamma$ & $x_{min}$ & $p$-value \\ 
		\noalign{\smallskip}\hline\noalign{\smallskip}
		Drug community	  &  $16553$	   & $61021$       & $32.08$ ($9.20$) & $160$            &  $1.57$ & $19$      & $.97$       \\ 
		Susceptible		  &  $3586$	   	   & $16499$       & $32.14$ ($8.75$) & $72$             &  $1.66$ & $8$       & $.84$       \\ 
		Immune			  &  $78463$	   & $496018$      & $30.31$ ($8.03$) & $323$            &  $1.51$ & $10$      &  $.76$          \\[5pt]
		\emph{Total}	  &  $98602$	   & $982197$      & $30.71$ ($8.32$) & $524$            &  $1.54$ & $8$       & $.57$       \\ 	
		\noalign{\smallskip}\hline
	\end{tabular}
\end{table*}

\section{Conclusions/Discussion}
\label{sec:5}

Drug abuse has seen a dramatic increase in the Russian Federation during the last two decades (Sunami, 2007; Mityagin, 2012)\nocite{Sunami2007,Mityagin2012}. The  rapid spread and extent of this `drug epidemic' forms a serious cause for alarm and finding effective ways to halt the current trend is of outmost importance. 

Due to the criminal nature and the general social disapproval of narcotics, it is difficult (or outright impossible) to assess the drug community directly. Official governmental statistics do provide some insight, but fail to give the complete picture; the `moderate' drug user is hardly noticed. Information retrieved from social networks such as LiveJournal can, therefore, contribute in gaining a better understanding of what constitutes the drug community in the Russian Federation and might prove to be vital for devising more effective intervention strategies. 

In this paper we present a method to assess this non-directly observable community by mining the popular social network site LiveJournal. By comparing the users' blogs with a dictionary consisting of known drug-related Russian terminology, we were able to identify those users that write most actively about drug use. By collecting their interests, we were able to create a general picture of the kind of users that can be found within the on-line drug community, see Table \ref{tab:3} and Fig. \ref{fig:3}. In addition, we introduced a naive Bayesian classifier for identifying potentially susceptible users by comparing their personal interests with the interests most common within the on-line drug community. The `infectious', `susceptible' and `immune' subnetworks were shown to have a similar structure; their degree distributions follow a power-law, although with slightly varying exponents. 

It is unclear to what extent we were able to identify the users that are really involved in drug use. Users that tend to write often about narcotics might do so for the following three reasons: 1) to raise the discussion on the social problems caused by drug abuse or propose possible ways to change the current situation for the better, 2) in an attempt to persuade others to stop or never start using drugs, i.e., `anti-propaganda', or 3) to share their experiences with drugs or to express their interest in this topic. We are solely interested in the group of users writing about narcotics for the third reason; they are the ones that use drugs or are likely to do so in the future. 

The appearance of the theme politics in Fig. \ref{fig:3} might be best explained by the presence of users in LiveJournal that do not write about drugs because they are personally interested or using them, but rather since they want to bring the social problems related to narcotics under the attention. The same might hold for the themes as the social and exact sciences, psychiatry and, potentially, nationalism. The presence of a theme like Christianity (consisting of the interests `God', `the Russian orthodox church' and `religion') is more likely to be explained by the presence of users that spread anti-propaganda, especially when taking the negative stance of the church towards drugs into account. 

Themes such as the occult, esotericism, science fiction and eastern teachings, however, are hardly explained by stating that the users interested in these topics are heavily concerned with the social impact of drug abuse, or actively spreading anti-propaganda.  Most likely, we caught a glimpse of the actual drug community. 

The explanations of why certain themes are presented in the on-line drug community are, of course, based solely on the view of the authors and, therefore, subjective. 
Further research is required to establish what themes are truly related to the Russian drug community. In order to establish the validity of the approach described in this paper, one might compare the presented results with law enforcement data, e.g., it would be interesting to compare the number of convictions for drug-related crimes between the on-line drug community and the rest of the crawled LiveJournal population.  

The susceptibility of an individual to drugs was determined on the basis of the similarity between his/her personal interests and the interests common in the on-line drug community. We limited ourselves here to their interests, since it was unclear how to relate the susceptibility of a user and his/her texts. 

The number of susceptible users is relatively small due to the small number of common interests present in the LiveJournal network. In fact, it can be shown that the frequency with which a certain interest occurs follows a power-law with exponent $\gamma \approx 1.54$, see Appendix 1. With a low number of common interests, there is often not enough to go on to identify a user as being susceptible. It is, thus, very well possible that we overlooked several immune users who should have been noted as being susceptible. 

Users were considered to be a member of the on-line drug community when the weighted sum of their blog entries crossed the threshold of 8, see Fig. \ref{fig:1}. We experimented with different thresholds and found that, although the list of significant interests does vary, the resulting clusters/themes remain stable. The weights assigned to the official and informal/`slang' terminology in the drug-dictionary were not varied. Since the final themes did not vary much while varying the threshold, it is unlikely that they would now. 

As mentioned before, we found that the LiveJournal network and the infectious, susceptible and immune subnetworks are most likely scale-free (i.e., their degree distributions follow a power-law). Although the performed goodness of fit test (Clauset et al., 2009)\nocite{Clauset2009} does not exclude other possibilities, e.g., Poisson, we can state with certainty that the distributions are heavy-right tailed, which entails that the network has hubs, i.e., users with a far higher degree than the rest of the network. This knowledge might be of major importance when one wants to disrupt the network to, for example, limit the spread of drug-related information on the network. Removing the hubs would heavily disrupt the information flow (Bollobas and Riordan, 2004; Albert et al., 2000; Crucitti et al., 2003)\nocite{Bollobas2004, Albert2000, Crucitti2003}.

This paper has shown the promise of `crawling social networks' in delineating and analyzing social groups that hitherto have eluded such research, because of the fundamentally opaque nature of membership of such groups. The case in point is the Russian drugs community. We hope that continuing research along the lines we set out in this paper will help to map the dynamics of this group, and will ultimately contribute to halting, if not reverting its tragic trend to grow.

\begin{acknowledgements}
The authors thank Dr. Sergey Mityagin from the Saint Petersburg Information and Analytical Center (SPb IAC) for fruitful discussions on the drug addiction profiles in the Russian Federation. In addition, the authors would like to express their gratitude to Prof. dr. T.K. Dijkstra from the University of Groningen (RUG) and the Free University Amsterdam (VU) for introducing us with false discovery rate control and his useful remarks. This work is supported by the \emph{Leading Scientist Program} of the Russian Federation, contract 11.G34.31.0019, as well as  by the Complexity program of NTU, Singapore. Peter Sloot also acknowledges the support from the FET-Proactive grant TOPDRIM, number FP7-ICT-318121.
\end{acknowledgements}

\section*{Appendix 1: LiveJournal User Interests}

\begin{figure*}[htb]
        \centering
        \begin{subfigure}[b]{0.544\textwidth}
                \centering
                \includegraphics[width=\textwidth]{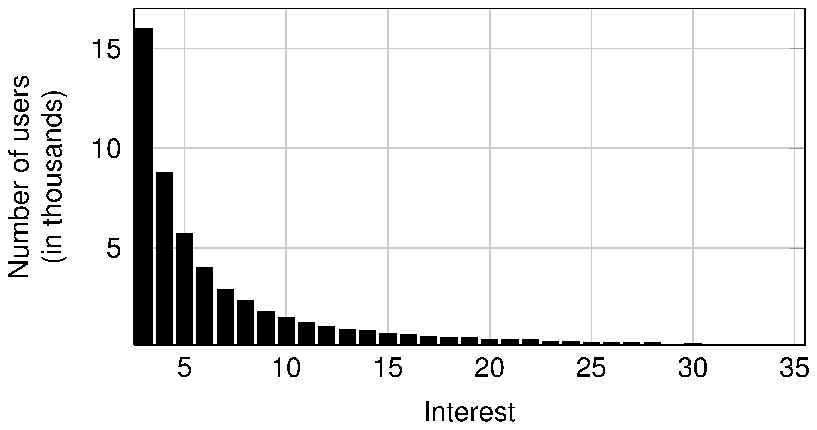}
                \caption{}
                \label{fig:6a}
        \end{subfigure}
        ~
        \begin{subfigure}[b]{0.232\textwidth}
               	\centering
               	\includegraphics[width=\textwidth]{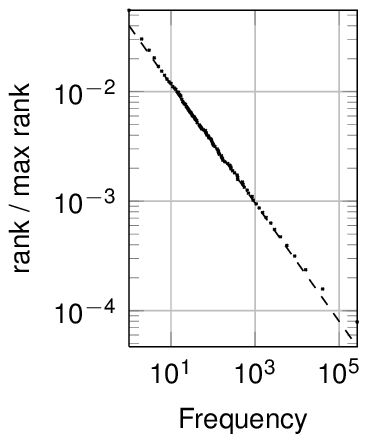}
               	\caption{}
               	\label{fig:6b}
        \end{subfigure}
        \caption{\textbf{a} The histogram of interests expressed by the users in the crawled LiveJournal data set \textbf{b} The rank/frequency log-log plot of the histogram in 6a and the maximum likelihood power-law fit ($\gamma \approx 1.54$ and $x_{min} = 3$)}
        \label{fig:6}
\end{figure*}

In this appendix we take a closer look at the frequency with which interests are expressed by the users of the social network LiveJournal. Fig. \ref{fig:6} shows the frequency of occurrence of interests within the crawled population. Note that the distribution is heavy right-tailed; its slope suggests that the distribution might follow a power-law, see eq. (\ref{eq:5}). Fig. \ref{fig:6}b shows the corresponding rank/frequency log-log plot of the histogram in \ref{fig:6}a. The exponent $\gamma \approx 1.54$ and the start of the distribution $x_{min} = 3$ were approximated using the maximum likelihood method as proposed by Clauset et al. (2009)\nocite{Clauset2009}. Note that the fitted line in \ref{fig:6}b approximates the distribution quite well. The standard goodness-of-fit test (Clauset et al. 2009)\nocite{Clauset2009} indicates there is no reason to believe that the distribution does not follow a power-law, i.e., the $p$-value was approximately equal to $.57$.

The fact that the distribution of interests within the SNS LiveJournal is heavy-right tailed explains why the number of susceptible users (see Table \ref{tab:4}) is relatively small compared to the other groups.

\end{document}